\documentclass[preprint,amsmath,amssymb,aps,physrev]{revtex4}

\usepackage{graphicx}
\usepackage{dcolumn}
\usepackage{bm} 
\usepackage{color}

\newcommand{\w}[1]{{\tilde{#1}}}
\renewcommand{\t}{\theta}

\newcommand{\df}{\dfrac}

\newcommand{\cA}{{\cal{A}}}
\newcommand{\cB}{{\cal{B}}}

\begin{document}

\title{Delayed transitions, promoted states and multistability in a pressure-driven nematic under an electric field}

\author{G.~McKay}
\affiliation{Department of Mathematics and Statistics,
  University of Strathclyde, Glasgow G1 1XH, UK}

\author{N.J.~Mottram}
\affiliation{School of Mathematics and Statistics, University of Glasgow, Glasgow G12 8QQ, UK}

\date{\today}

\begin{abstract} 
We consider the effects of an applied pressure gradient on the classical Freedericksz transition, finding a delayed transition, the promotion of particular director configurations and even pressure-induced multistability. Using the theoretical framework developed by Ericksen and Leslie, we find that the applied pressure gradient adapts the normal pitchfork bifurcation at critical applied voltage, leading to both a delayed bifurcation to higher voltages and a transformation from a supercritical to a subcritical bifurcation so that within a range of voltages there are at least two possible steady states. This range of voltages grows with increasing pressure gradient and eventually includes the zero voltage state so that, for sufficiently strong flow, there are at least two steady states at zero applied voltage. For sufficiently high pressure gradients, we also find that flow-alignment can create a completely new attracting steady state, one that is unstable without flow. We provide a flow-strength-electric field parameter plane that summarises the parameter ranges for which there are multiple steady states and suggest realistic mechanisms to move between these states, as well as an analytical model for the delayed Freedericksz transition effect. The novel steady states found in this work give the possibility of director and flow hysteresis in microfluidic devices.     
\end{abstract}


\maketitle

\section{Introduction}
In anisotropic liquids, the orientation of the anisotropic axis can be influenced through several mechanisms -- such as surface interactions, viscous stresses, and externally applied electric or magnetic fields. A key example of these materials is nematic liquid crystals, where the average molecular alignment, known as the director, both affects and is affected by flow dynamics and electric fields \cite{Miesowicz1936,Miesowicz1946, IWS1}. The intricate relationship between director orientation and fluid behavior plays a crucial role in liquid crystal display technology, where feedback between flow and alignment may pose significant challenges \cite{Lueder} and underpins emerging photonic and display devices, where droplet manipulation is essential, either in the manufacturing or operation \cite{HANDBOOK_ODF, Zhao2013, Xu2013, Cousins1, Cousins2}. The underlying physical origin of these effects is the director which, as a symmetry-breaking axis, results in an asymmetric stress tensor, anisotropic electrical polarisation and an intrinsic elasticity within the fluid  \cite{Oseen1933,Frank1958,Eriksen1960,Leslie1968,deGennes1995}. In this paper we focus on three core phenomena: (1) the application of an electric field can reorient the director via the Freedericksz transition (2) an applied pressure gradient leads to a shear stress that tends to align the director with a preferred angle relative to flow direction -- a behavior known as flow alignment \cite{Leslie1968}, that can potentially compete with field-induced alignment; and (3) as a consequence of the $\pi$-rotational symmetry of the director, the flow alignment, coupled with the field-induced alignment, may create a steady state that would not be accessible without flow. Although field-induced changes in rheology have been observed before
\cite{ Yang1992,Yoshida2002, Negita1996,DeVolder2006,Reyes2008,Patricio2012,Ananthaiah2014, MottramUTube}, here we concentrate on the opposite effect, flow-induced changes to the orienting field effects. Using the theoretical framework developed by Ericksen and Leslie \cite{Eriksen1960,Leslie1968}, we find that the applied pressure gradient adapts the classical pitchfork bifurcation at the Freedericksz critical applied voltage, leading to a subcritial, rather than a supercritical bifurcation, and at a higher voltage. This means that, within a range of voltages there are at least two possible steady states, and for sufficiently strong flow there are at least two steady states at zero applied voltage. 

\section{Model}
\begin{figure}
 \begin{center} 
 \resizebox{.8\textwidth}{!}{\includegraphics{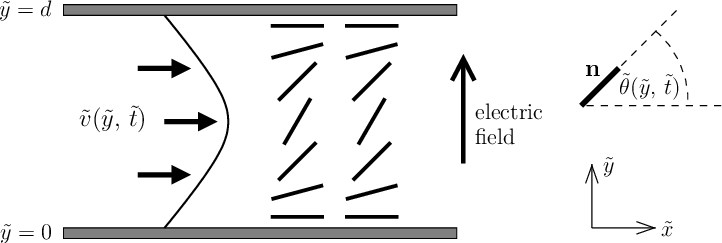}}\\
 \caption{\label{fig1}  Sketch of the nematic liquid crystal layer between substrates at $\w{y}=0$ and $\w{y}=d$. The director, which lies in the $\w{x}\w{y}$-plane at an angle $\w{\t}$ to the $\w{x}$-axis, is  strongly anchored at  the substrates so that  $\w{\t}(0)=\w{\t}(d)=0$.  A pressure gradient in the $\w{x}$-direction induces flow with velocity $\w{v}$. Both the director angle and flow speed depend on the $\w{y}$ coordinate and time $\w{t}$.}
 \end{center}
 \end{figure}

We consider a nematic liquid crystal layer of thickness $d$ confined between parallel substrates at $\w{y}=0$ and $\w{y}=d$ (see Fig.~\ref{fig1}). The substrates include electrodes, allowing an electric field to be applied in the $\w{y}$ direction, and are treated so that the director at the substrates (the local average molecular long-axis direction) aligns with the $\w{x}$-direction. A pressure gradient in the  $\w{x}$-direction leads to flow with a horizontal velocity,  $\w{v}(\w{y},\,\w{t})$. 
With both an applied pressure gradient and an applied voltage, the director experiences three competing torques: the reorienting effects of the shear gradient which attempt to align the director to a Leslie angle, a function of two of the Leslie viscosities; the reorientation effects due to the electric field which, since we assume a positive dielectric nematic, attempt to align the director in the $\w{y}$-direction; and an elastic torque which attempts to create a uniform director orientation in the $\w{x}$-direction due to the substrate alignment. As we will see below, these three competing effects produce multiple attracting director orientations and a rich bifurcation structure. 

Using a standard approach, the nematic director, ${\mathbf{n}}$, is assumed to lie in the  $\w{x}\w{y}$-plane, so that ${\mathbf{n}}=(\cos\w{\t},\, \sin\w{\t},\,   0 \bigr)$, where $\w\theta(\w{y},\,\w{t})$ is the director tilt angle with respect to the $\w{x}$-direction, and the flow is assumed rectilinear in the $\w{x}$-direction with velocity $\w{v}(\w{y},\w{t})$. The director angle and flow  velocity can then be modelled using the Ericksen-Leslie  
equations for nematics~\cite{Eriksen1960,Leslie1968}. For this form of the director and flow, the governing equations for $\w{\t}$ and $\w{v}$ are 
\begin{align}
\hspace*{-1cm}{\gamma}_1 \w{\t}_\w{t}&= (K_1\cos^2\w{\t}+K_3\sin^2\w{\t})\w{\t}_{yy} +\df{(K_3-K_1)}{2}({\w{\t}_{y}})^2\sin2\w{\t}-\w{m}(\w{\t}) \w{v}_\w{y}  +\df{\Delta \epsilon\, \epsilon_0}{2}(\w{U}_\w{y})^2\sin2\w{\t},\label{EL1A}\\
\rho \w{v}_\w{t} &=-\frac{\Delta p}{L}+ \bigl(\w{g}(\w{\t}) \w{v}_{\w{y}} + \w{m}(\w{\t})\w{\t}_\w{t}\bigr)_\w{y},\label{EL2A}
\end{align}
where
\begin{align}\label{ELmg}
\w{m}(\w{\t})&= \alpha_3\cos^2\w{\t}-\alpha_2\sin^2\w{\t},\\
\w{g}(\w{\t})&=
\frac{1}{2}\bigl(\alpha_4+(\alpha_5-\alpha_2)\sin^2\w{\t}+(\alpha_3+\alpha_6)\cos^2\w{\t}\bigr)+\alpha_1\sin^2\w{\t}\cos^2\w{\t},  \label{ELmg2}
\end{align}
 with $\alpha_i$ ($i=1,\ldots, 6$) the Leslie coefficients for the nematic. The rotational viscosity ${\gamma}_1=\alpha_3-\alpha_2$, and $\epsilon_0$ is the permittivity of free space. The dielectric anisotropy,  $\Delta\epsilon=\epsilon_{||}-\epsilon_\perp$, is the difference between
 the dielectric permittivities of the nematic parallel and perpendicular to the director.
  Subscripts $\w{t}$ and $\w{y}$ represent partial derivatives with respect to the variable. 
 Coefficients $K_1$ and $K_3$ are the elastic constants associated with, respectively, splay and bend of the nematic director configuration, $\rho$ is the fluid density and $\Delta p/L$ is the applied pressure gradient, e.g.~a pressure difference $\Delta p$ applied between two points a horizontal distance $L$ apart. 
The electric potential $\w{U}(\w{y},\,\w{t})$ is governed by Maxwell's equations for a dielectric material with no free
charge~\cite{Kaiser}, 
\begin{equation}
\bigl((\epsilon_{||}\sin^2\w{\t}+\epsilon_\perp\cos^2\w{\t})\w{U}_{\w{y}}\bigr)_\w{y}=0.\label{max}
\end{equation}
As mentioned above, the substrates are treated so that the director is forced to lie in the $\w{x}$-direction, so-called infinite anchoring. We assume non-slip conditions on the flow speed and impose a potential difference $V$ across the cell, so that
\begin{equation}
\w{\t}(0,\,\w{t})= \w{\t}(d,\,\w{t})=0,\quad 
\w{v}(0,\,\w{t})= \w{v}(d,\,\w{t})=0,\quad
\w{U}(0,\,\w{t})=0,\quad  \w{U}(d,\,\w{t})=V.\label{BC3}
\end{equation} 

We now non-dimensionalise equations (\ref{EL1A})--(\ref{BC3}) by introducing the scalings
\begin{equation}
\w{y}= d\, {y}, \quad  \w{t} = T\,{t}, \quad \w{U}(\w{y},\w{t})= V\,{U}(y,t),\quad \w{v}(\w{y},\w{t})=  W\,{v}(y,t),\label{nondim}
\end{equation}
where $T$ and $W$ represent characteristic time and velocity scales for our system, which will be determined below by considering the relevant physical effects in the model. The director angle is not rescaled by the non-dimensionalisation, but we remove the tilde to define the director angle, $\t$, as a function of the non-dimensional variables $y$ and $t$.  The viscosity function $\w{m}$ is also normalised using the appropriate viscosity, namely,  the rotational viscosity $\gamma_1$, while the function $\w{g}$ is normalised using the Newtonian viscosity $\mu=\alpha_4/2$,   
\begin{equation}\label{normg}
\w{m}({\w{\t}})=  \, {\gamma}_1 {m}({\t}),\quad \w{g}(\w{\t} ) =  \mu {g}({\t}).
\end{equation}

Given that we are interested in the reorientation of the director, we now choose the timescale to be the classic timescale of elastic reorientation of the director. Since elastic- and voltage-induced flow is often transient, we will choose the velocity scale to be that of a pressure-driven system. The characteristic scalings $T$ and $W$ are then
\begin{equation}
\label{nondim2} T=\df{\gamma_1 d^2}{K_1},\quad W= -\df{d^2 \Delta p}{\mu L},
\end{equation}
and equations  (\ref{EL1A}), (\ref{EL2A}) and (\ref{max})  become
\begin{align}
\t_{t}&= (\cos^2\t+k\sin^2\t)\t_{yy} +\df{(k-1)}{2}({\t_{y}})^2\sin2\t-{{\cal E}} \, {m}(\t)\,  {v}_y  +\df{\pi^2 {\bar{V}}^2}{2} 
 (U_y)^2\sin2\t,\label{EL1ND}\\[5pt]
\df{{\cal R}}{{\cal E}} 
{v}_{t} &= 1 
+ \left( {g}(\t) {v}_{y} + \df{\nu}{{\cal E}} 
 {m}(\t)\t_{t}\right)_y,\label{EL2ND}\\[5pt]
0&=\left((\epsilon\sin^2\t+\cos^2\t)U_y\right)_y,\label{maxND}
\end{align}
where we now have the non-dimensional parameter  $\bar{V}=V/V_c$, the voltage normalised with respect to the critical voltage for the classic Freedericksz transition $V_c=\pi\sqrt{K_1/(\epsilon_0\Delta \epsilon )}$~\cite{IWS1}. The other non-dimensional parameters are   $k=K_3/K_1$, a measure of elastic anisotropy;  $\nu=\gamma_1/\mu$, the ratio of rotational and Newtonian shear viscosities;   $\epsilon=\epsilon_{||}/\epsilon_{\perp}$, a measure of the dielectric anisotropy; and 
\begin{equation}
{{\cal E}} =  \dfrac{\gamma_1 Wd}{K_1}=-\df{\gamma_1 \Delta p\, d^3}{\mu K_1\,  L},\qquad {{\cal R}}=\dfrac{\rho W d}{\mu}= -\df{ \rho\, \Delta p\,d^3}{\mu^2 L},
\end{equation}
which are, respectively, the Ericksen and Reynolds numbers for this pressure-driven system~\cite{CALD1,CALD2}. The Ericksen number was first introduced by de Gennes~\cite{deG} and is a measure of the ratio of viscous to elastic torques on the director, while the Reynolds number is a ratio of the viscous to inertial forces on a fluid element. Typically, for the flow of nematic liquid crystals the Reynolds number is small while the Ericksen number is relatively large with ${{\cal R}}/{{\cal E}}\sim 10^{-4}$~(see Stewart~\cite{IWS1}). Therefore, it is common to discard the fluid inertia term on the left hand side of  (\ref{EL2ND}). 
Finally, with the scalings introduced above, the  boundary conditions (\ref{BC3}) can now be written as
\begin{equation}
\t(0,\,t)= \t(1,\,t)=0,\quad 
v(0,\,t)=v(1,\,t)=0,\quad U(0,\,t)=0,\quad  U(1,\,t)=1.\label{BC3ND}
\end{equation}

Once the system reaches steady state, the director angle and flow speed must satisfy steady-state versions of (\ref{EL1ND}) and (\ref{EL2ND}),
\begin{align}
0&= (\cos^2\t+k\sin^2\t)\t_{yy} +\dfrac{(k-1)}{2}({\t_{y}})^2\sin2\t -{{\cal E}} \, {m}(\t)\,  {v}_y  +\dfrac{\pi^2 {\bar{V}}^2}{2} 
 (U_y)^2\sin2\t,\label{SS1}\\[5pt]
 0&= 1 
+ \bigl( {g}(\t) {v}_{y} \bigr)_y\,,\label{SS2}
\end{align}
with the potential $U$ still determined by (\ref{maxND}).

\section{Numerical Results}
\begin{figure}
 \begin{center} 
 \resizebox{.75\textwidth}{!}{\includegraphics{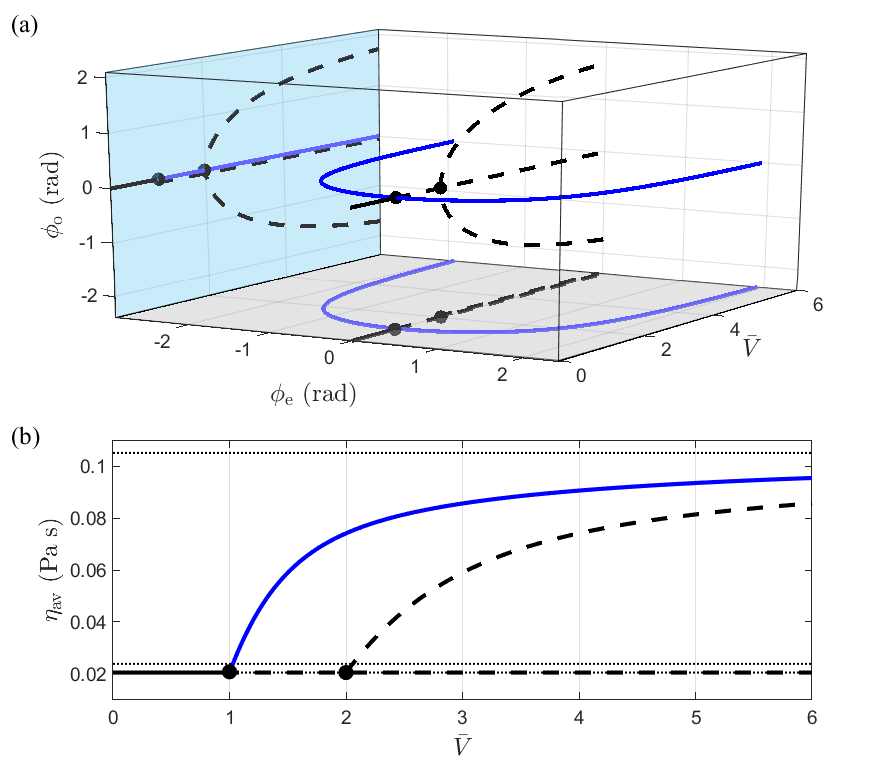}}
 \caption{\label{fig2} Dependency of  (a) director angle measures $\phi_{\mathrm{e}}$, $\phi_{\mathrm{o}}$  and (b) effective viscosity $\eta_{\mathrm{av}}$  upon the normalised voltage $\bar{V}$ for Ericksen number ${{\cal E}}=0$. 
 The curves represent attracting (solid lines) and non-attracting (dashed lines) steady states  for  the first three modes: $n=0,\, 2$ (black); $n=1$ (blue).  The horizontal dotted lines in (b) indicate $\eta_1=0.0204$, $\eta_L=0.0237$ and $\eta_2=0.1052$.}
\end{center}
 \end{figure}

\begin{figure}
 \begin{center} 
 \resizebox{\textwidth}{!}{\includegraphics{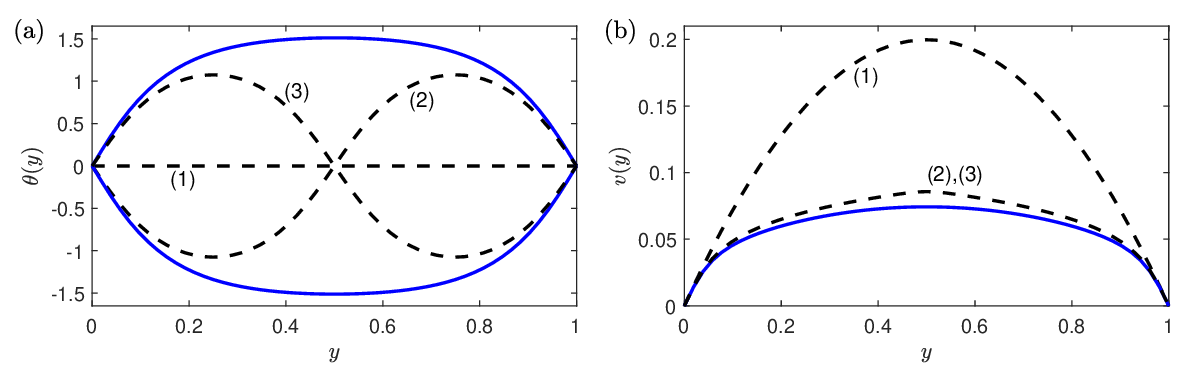}}\\
 \caption{\label{fig3} Steady states solutions for (a) director angle $\theta(y)$ and (b) velocity $v(y)$  when ${{\cal E}}=0$ and $ \bar{V}=3$. The curves represent attracting (solid lines) and non-attracting (dashed lines) steady states  for the  modes $n=0,\,2$ (black) and $n=1$ (blue). Velocities corresponding to each unsteady director profile indicated by (1), (2) and (3). }
 \end{center}
 \end{figure}

\begin{figure}
 \begin{center} 
 \resizebox{.75\textwidth}{!}{\includegraphics{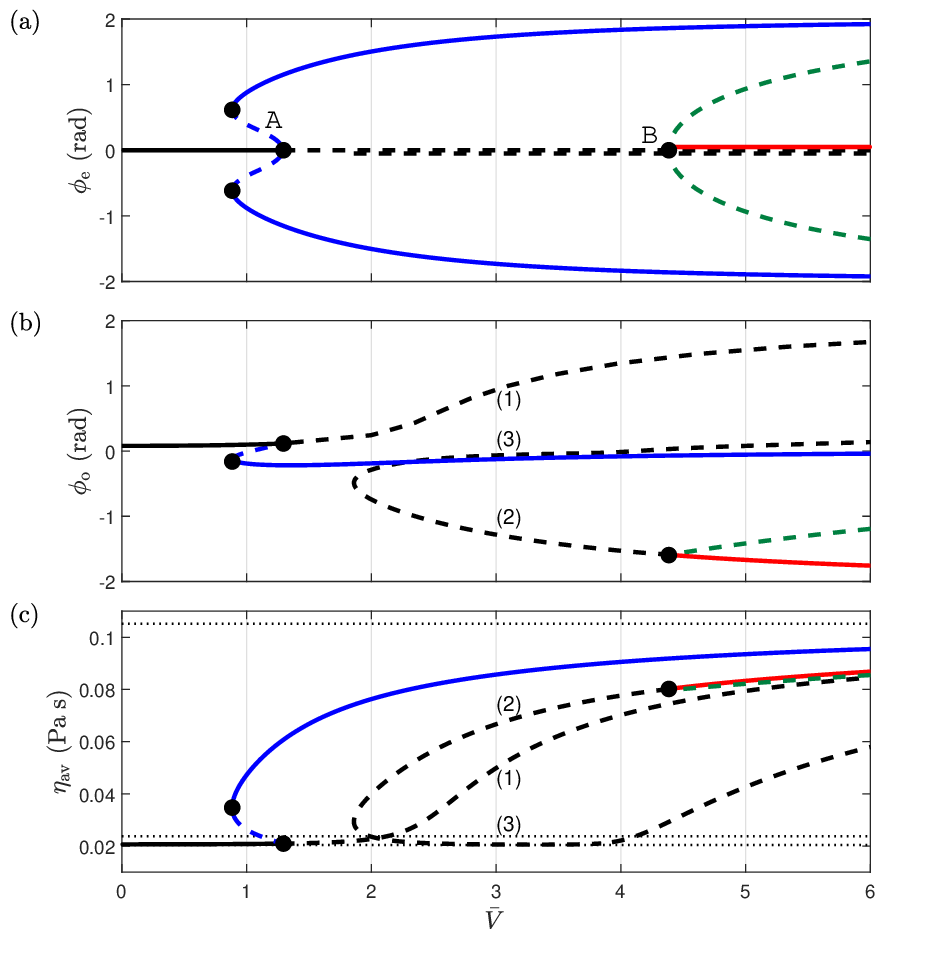}}
 \caption{\label{fig4} Dependency of    $\phi_{\mathrm{e}}$, $\phi_{\mathrm{o}}$  and   $\eta_{\mathrm{av}}$  upon the normalised voltage $\bar{V}$ for Ericksen number ${{\cal E}}=150$ obtained by continuation of the first three modes. 
 The curves represent attracting (solid lines) and non-attracting (dashed lines) steady states, with attracting steady states (red) and secondary bifurcation (green) at higher voltages. The (1), (2), (3) numbered points on the branches correspond to the numbering in Fig.~\ref{fig3}.} 
\end{center}
 \end{figure}

\begin{figure}
 \begin{center} 
 \resizebox{\textwidth}{!}{\includegraphics{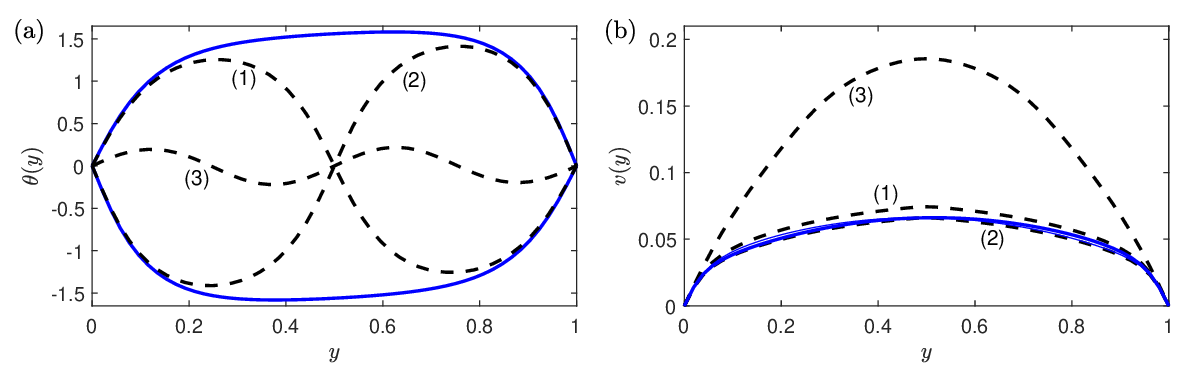}}\\
 \caption{\label{fig5} Steady states solutions for (a) director angle $\theta(y)$ and (b) velocity $v(y)$  when ${{\cal E}}=150$ and $ \bar{V}=3$ obtained by continuation of the first three modes. The curves represent attracting (solid lines) and non-attracting (dashed lines) steady states, with velocities corresponding to each unsteady director profile indicated by (1), (2) and (3).  } 
 \end{center}
 \end{figure}

\begin{figure}
 \begin{center} 
 \resizebox{.75\textwidth}{!}{\includegraphics{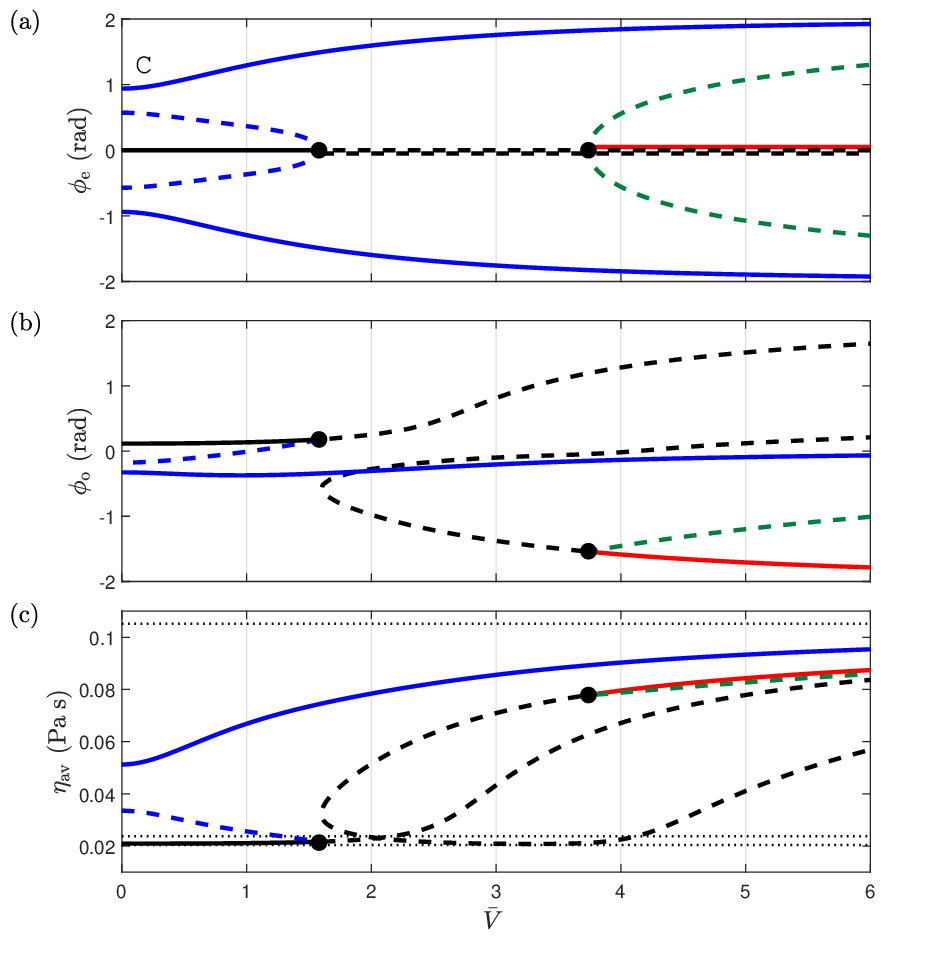}}
\caption{\label{fig6}Dependency of  $\phi_{\mathrm{e}}$, $\phi_{\mathrm{o}}$  and   $\eta_{\mathrm{av}}$  upon the normalised voltage $\bar{V}$ for Ericksen number ${{\cal E}}=250$ obtained by continuation of the first three modes. 
 The curves represent attracting (solid lines) and non-attracting (dashed lines) steady states, with attracting steady states (red) and secondary bifurcation (green) at higher voltages.} 
\end{center}
 \end{figure}

\begin{figure}
 \begin{center} 
 \resizebox{.85\textwidth}{!}
 {\includegraphics{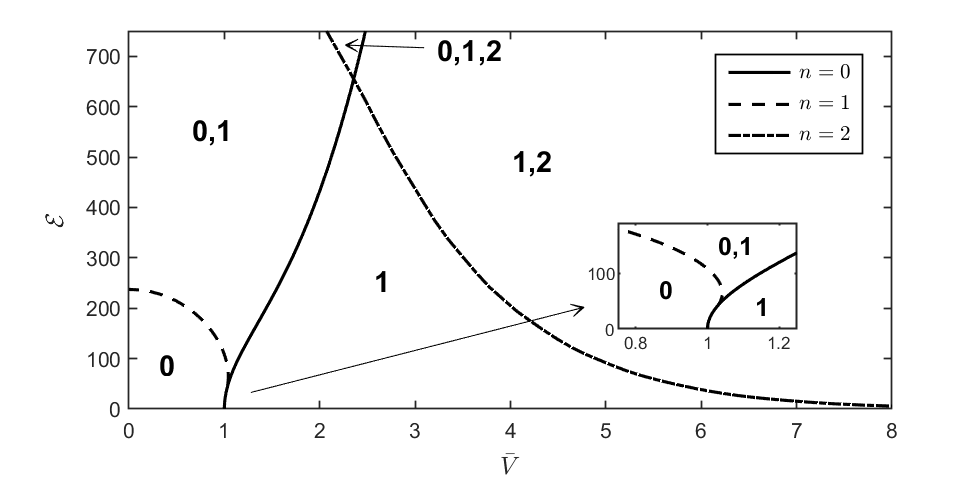}}\\
 \caption{\label{fig7}
Regions in the normalised voltage-Ericksen number space, $(\bar{V},\,{{\cal E}})$, corresponding to whether $n=0,\,1$ and 2 modes are attracting steady states.
} 
 \end{center}
 \end{figure}

\begin{figure}
 \begin{center} 
 \resizebox{.8\textwidth}{!}{\includegraphics{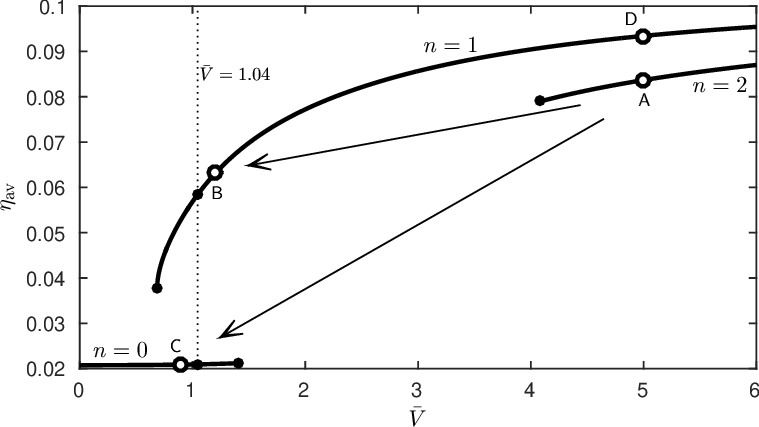}}
 \caption{\label{fig8}The effective viscosity $\eta_{\mathrm{av}}$ of attracting steady states when ${{\cal E}}=190$, versus normalised voltage. The dotted line at $\bar{V}=1.04$ indicates the transition where $n=0$ or $n=1$ is the preferred state upon a sudden reduction of voltage from point \textsf{A}.
 } 
 \end{center}
 \end{figure}

In Figs~\ref{fig2} to \ref{fig9}, we consider the nematic liquid crystal 5CB at 26$^\circ$C, with the Leslie viscosities and other material properties as provided by Stewart~\cite{IWS1}. For this material, the critical Freedericksz transition voltage is $V_c=0.78\,$V, while the elastic anisotropy ratio $k=0.63$. The dielectric anisotropy is positive, so the electric field acts to align the director perpendicular to the direction of flow.
Equations (\ref{maxND})--(\ref{SS1}) allow us to calculate the steady-state director profile for a given normalised voltage $\bar{V}$ and Ericksen number ${\cal E}$. In order to examine the behaviour of our system over a range of $\bar{V}$ and ${\cal E}$, we use two measures to represent a steady solution $\t(y)$.
Commonly, the maximum distortion angle
$ \max_{y\in[0,1]} |\theta(y)|$ is employed to  characterise  steady states in analyses of Freedericksz transitions. However, in our context, this measure often fails to distinguish between multiple steady solutions and provides no information about the symmetry of the solution (e.g., if $\t(y)$ is symmetric or antisymmetric about the centre of the layer). 
Instead, we use two measures of the director angle  in order to characterise its behaviour, 
\begin{equation}
\phi_{\rm e}=\frac{\langle\,\theta(y)\,\theta_{\rm 1}(y)\, \rangle}{
\langle\, \theta_{\rm 1}(y)^2\, \rangle},\qquad 
\phi_{\rm o}=
\frac{\langle\,\theta(y)\,\theta_{\rm 2}(y) \,\rangle}{ \langle\, 
\theta_{\rm 2}(y)^2\, \rangle},  
\label{meas}
\end{equation}
where $\langle\,\cdot\,\rangle$ represents integration across the channel from 
$y=0$ to $1$, and
\begin{equation}\theta_{n}(y) =\sin(n\pi y) , \qquad  n\in\mathbb{N}_0 
\label{modes}
\end{equation}
Here, the subscripts e and o refer to, respectively, even and odd (or symmetric and antisymmetric) contributions with respect to the centre of the cell. 
For a steady state, ${g}\bigl(\theta(y)\bigr)$  is a position-dependent, non-dimensional, effective viscosity. From this, we can derive another measure to characterise the steady state, an average effective viscosity,
\begin{equation}
 \eta_{\mathrm{av}}=\mu \int_0^1 {g}\bigl(\theta(y)\bigr)\,\mathrm{d}y,\label{gav}
\end{equation}
where $\mu$ is the Newtonian viscosity, included so that $\eta_{\mathrm{av}}$ is the dimensional effective viscosity. 

Figures~\ref{fig2}, \ref{fig4} and \ref{fig6}
present $\phi_{\mathrm{e}}$, $\phi_\mathrm{o}$  and $\eta_{\mathrm{av}}$ for steady-state director profiles over a range of normalised voltages  and selected Ericksen numbers. 
The three-dimensional equilibria branches in Fig.~\ref{fig2}(a)  have been projected onto the shaded horizontal and vertical planes  in order to help visualise the symmetric  and antisymmetric components of each branch. Due to the complexity of the plots, the three-dimensional structure has been replaced in Figs~\ref{fig4} and~\ref{fig6} by the individual $\phi_\mathrm{e}$ and $\phi_\mathrm{o}$ projections. 

In the viscosity subplots in Figs~\ref{fig2}, \ref{fig4} and \ref{fig6}, we have also provided lines indicating the Miesowicz viscosities $\eta_1=(\alpha_3+\alpha_4+\alpha_6)/2=0.0204$ and $\eta_2=(-\alpha_2+\alpha_4+\alpha_5)/2=0.1052$ for the material under consideration. These are the limiting values of the  viscosity function $\w{g}$ defined in (\ref{ELmg2}), corresponding to the cases where ${\theta}\equiv 0$ and ${\theta}\equiv\pi/2$, respectively.  Finally, the flow alignment angle, or Leslie angle, for a nematic $\theta_L$ is defined by $\tan^2\theta_L=\alpha_3/\alpha_2$~\cite{IWS1}. From this we can calculate  an equivalent effective viscosity, $\eta_L=\w{g}(\theta_L)=0.0237$, which is also shown in the subplots. 
Figures \ref{fig3} and \ref{fig5} 
provide steady-state solutions for the director angle $\theta(y)$ and velocity $v(y)$ at specific voltages for the same Ericksen numbers considered in Figs~\ref{fig2} and \ref{fig4}, respectively. 
All calculations have been carried out in \texttt{MATLAB} using the continuation package \texttt{MATCONT}~\cite{MATCONT}.

In Figs~\ref{fig2} and \ref{fig3}  when there is no applied pressure gradient so  ${{\cal E}}=0$, we observe the classic Freedericksz transition behaviour. This behaviour is well known and described here in order to contrast with the system  when the applied pressure gradient is non-zero. For no applied voltage, the trivial state $\t\equiv 0$, which we call the $n=0$ mode, is the solution and, as the voltage increases, instabilities to eigenmodes of the form $\theta_n(y)=\sin(n\pi y)$ occur at $\bar{V}=1,\,2\,\ldots$, which we call the $n=1,\,2,\,\ldots$ modes. 
Figures~\ref{fig2} and \ref{fig3} are restricted to the modes $n=0$, $1$ and $2$, with Fig.~\ref{fig3} illustrating steady director angle profiles $\theta(y)$, and corresponding  flow velocity  $v(y)$, when ${\cal{E}} = 0$ and $\bar{V}=3$.
The trivial state, the $n=0$ mode, first becomes  
non-attracting at  $\bar{V}=1$ in Fig.~\ref{fig2}. For  voltages 
above this transition point, the system is attracted to symmetric $n=1$ steady solutions  given approximately by $\theta(y)\approx\theta_m\sin(\pi y)$, where $\t_\mathrm{m}$ is the maximum director angle. 
Higher mode instabilities from the $n=0$ state become possible for higher voltages but are not attracting steady states. For instance, for $\bar{V}>2$ the $n=0$ mode is non-attracting and the system will move away from the trivial state if it is perturbed by an antisymmetric $n=2$ mode. 
However, the $n=2$ mode itself is  non-attracting and symmetric perturbations 
will lead to the system being attracted to a solution on the $n=1$ branch. 

As the voltage increases, the magnitude of the maximum director distortion 
 approaches $\pi/2$ for the $n=1$ solutions 
as the director aligns with the electric field. 
 The two steady $n=1$ solutions are  equal in magnitude and opposite in sign in Fig.~\ref{fig3}(a) but, since $g$ is an even function, the effective viscosities of these solutions are equal in Fig.~\ref{fig2}(b).   The two $n=2$ solutions  also correspond to equal effective viscosities.
From Fig.~\ref{fig2}(b) we also see that, as expected,  the effective viscosity of the $n=0$ state is exactly $\eta_{\rm av}=\eta_1$. On the $n=1$ branch, the average effective viscosity $\eta_\mathrm{av}$ changes from $\eta_1$ when $\bar{V}\approx 1$, and approaches $\eta_2$ for large  $\bar{V}$  due to the magnitude of the maximum director distortion  increasing from zero towards $\pi/2$.

Figure~\ref{fig4}  shows the steady states derived by numerical continuation of the  branches in Fig.~\ref{fig2} when the Ericksen number is increased to   ${{\cal E}}=150$, with plots of $\theta(y)$ and $v(y)$ when $\bar{V}=3$ presented in Fig.~\ref{fig5}. The unsteady solutions  in  Fig.~\ref{fig5} correspond to different branches in Fig.~\ref{fig4}, as indicated by (1), (2) and (3). 
For non-zero Ericksen numbers, the addition of a pressure gradient   causes symmetry breaking. The $n=0$ and $n=2$ modes remain anti-symmetric with respect to the centre of the cell ($\phi_\mathrm{e}=0$) when ${{\cal E}}>0$, however the $n=1$ mode is no longer perfectly symmetric leading to $\phi_{\mathrm{o}}\not\equiv 0$.
The two $n=1$ solutions in Fig.~\ref{fig5}(a) also coincide with different velocity profiles in Fig.~\ref{fig5}(b).

In Fig.~\ref{fig4} we see that  the attracting $n=1$ branch has detached from the $n=0$ branch and they are now connected via non-attracting states, indicated by {A} in Fig.~\ref{fig4}(a). The supercritical pitchfork bifurcation at $\bar{V}=1$ has become a subcritical pitchfork bifurcation. This change in the bifurcation structure of the classical Freedericksz transition means that the trivial $n=0$ state now remains an attracting steady state beyond $\bar{V}=1$ for ${{\cal E}}>0$, indicating a delay in the Freedericksz transition due to the fluid flow. Similarly, the $n=1$ branches are now attracting steady states for voltages below the critical Freedericksz transition voltage of $\bar{V}=1$. The system is, therefore, exhibiting multiple attracting steady states, corresponding to the $n=0$ and $n=1$ modes, close to $\bar{V}=1$.

For ${{\cal E}}>0$ and for high voltages, one of the $n=2$ branches contains attracting steady states. This transition   coincides at a secondary bifurcation, indicated by {B} in Fig.~\ref{fig4}(a), with newly formed asymmetric steady states for which both $\phi_\mathrm{o}$ and $\phi_\mathrm{e}$ are non-zero. Although the average effective viscosity for these non-attracting secondary states is difficult to observe in Fig.~\ref{fig4}(c) as the values are close to those of the attracting $n=2$ branch, the $\phi_\mathrm{o}$ and $\phi_\mathrm{e}$ profiles are clear in Fig.~\ref{fig4}(a),(b).

For even larger Ericksen numbers, as shown in Fig.~\ref{fig6} 
for ${\cal{E}}=250$, the $n=0$ branch extends and delays the Freedericksz transition to even higher voltages, meaning that the flow-induced torque on the director is dominating electric field effects except at very high voltages. 
The  $n=1$ branches have connected with their mirror negative voltage states at $\bar{V}=0$, indicated by {C} in Fig.~\ref{fig6}(a),  so  multiple attracting steady states are now  possible  for zero applied voltage. 
In addition, the flow continues to stabilise one of the $n=2$ branches and, in a manner similar to the $n=1$ case in Fig.~\ref{fig4}, the  attracting $n=2$ branch extends to a wider range of voltages.  

To summarise the above results for a range of Ericksen numbers, Figure~\ref{fig7} illustrates the demarcation of $(\bar{V},\,{{\cal E}})$ space into regions of different levels of multistability. In this figure, the notation $0,\,1,\,2$ indicates regions in which the respective modes are attracting steady states. 
The $n=0$ region shows that with increasing applied pressure gradient (i.e.~increasing Ericksen number ${\cal E}$) the Freedericksz transition is delayed to higher voltages, as is also evident from  Figs~\ref{fig2}--\ref{fig6}. 
For sufficiently high Ericksen number (${\cal E}\gtrsim 220$), the $n=1$ mode is attracting for all voltages,
while the range of voltages over which the $n=2$ state is attracting increases as Ericksen number increases.  
There are two different parameter regimes of two-mode multistability with the $n=1$ mode coexisting with either the 0 or 2 state. A three-mode multistable region also exists for low voltages and high Ericksen numbers, i.e.~high flow rates. Point $(2.4,\,655)$ is a critical transition point between between regions of  one-, two- and three-mode multistability. The inset highlights the cusp in the $n=1$ transition curve and the delayed Freedericksz transition for the $n=0$ mode.

It should be noted that  in Figs \ref{fig2}--\ref{fig7} we concentrated on the first three modes $n=0,1,2$. There are higher modes, which can also become attracting steady states for sufficiently high Ericksen numbers. However, the qualitative behaviour of their regions of attraction in Fig.~\ref{fig7} are similar to those of $n=1,2$ and given the need for large applied pressure gradients, and their reduced areas of attraction in the $(\bar{V},\,{\cal E})$ diagram, we do not consider these states here.
 
In order to investigate the transient dynamic behaviour of this system, for instance when switching between branches, we have carried out the full numerical solution of the system in eqs~(\ref{EL1ND})--(\ref{maxND}) subject to boundary conditions (\ref{BC3ND}) using COMSOL Multiphysics~\cite{COMSOL}. In order to model the operation of a real system, we perturb our system through a rapid change in applied voltage. In this way, we can examine whether a sudden change in voltage leads to a transition between branches and whether the process is reversible. 

We can see from Fig.~\ref{fig7} that the $n=1$ branch of solutions 
is always present, except when both the applied voltage and Ericksen number are small, in which case only $n=0$ solutions exist. 
The details have been omitted here, but it is very clear from our dynamic model that, if it exists, it is the $n=1$ branch which acts the global energy minimiser in \emph{almost} all situations. 
Any perturbation of a $n=1$ state by a rapid change in voltage (increase or decrease) will result in a new solution on the same $n=1$ branch, provided it exists for the given Ericksen number and voltage.
Perturbing a $n=0$ or $n=2$ solution via a small voltage change may lead to a steady state on the same branch. However, for a significant change of voltage the system will normally transition to a $n=1$ solution if available.  

Perhaps the most interesting behaviour occurs for Ericksen numbers where $n=0$ and $n=1$ branches co-exist over a small range of voltages, and where the $n=2$ state is flow stabilised at high voltages. In Figure~\ref{fig8}, we consider the case ${{\cal E}}=190$ where the $n=0$ and $n=1$ branches coexist between $\bar{V}=0.7$ and $1.4$. This effective viscosity plot is similar to those in Figs~\ref{fig4} and \ref{fig6}, although we have omitted  the  non-attracting  branches in order to concentrate on attracting steady states. 
If we perturb any point on the $n=1$ branch by a sudden change in voltage, the final steady state will lie on the same branch, unless the final voltage $\bar{V}<0.7$ where $n=0$ is the only attracting branch of solutions. 
 For a small change in voltage on the flow-stabilised $n=2$  branch from, say, $\bar{V}=5.0$ (point \textsf{A}) to $4.5$, the resulting steady states are still type $n=2$. A reduction to a voltage between $\bar{V}=1.4$ and $4.1$ always leads to a transition to the only attracting branch, namely, $n=1$. In fact, this behaviour persists for voltages below $\bar{V}=1.4$ and greater than $\bar{V}=1.04$. However, for voltage $\bar{V}<1.04$ the $n=0$ branch becomes the preferred state. For example,  in Figure~\ref{fig8} an initial $n=2$ state at point \textsf{A} is transitioned to the $n=1$ state \textsf{B} by reducing the voltage from $\bar{V}=5.0$ to $1.2$, whereas reducing the voltage to $\bar{V}=0.9$  results in the  $n=0$ state at \textsf{C}. This transition from $n=2$ to $n=0$ or $n=1$, however, is not reversible. Increasing  the voltage back to $\bar{V}=5.0$ would see  both points \textsf{B} and \textsf{C}  evolving towards the steady state  \textsf{D} on the $n=1$ branch, not the original point \textsf{A}.

\section{Analysis of Freedericksz transition for small angle $\theta$}
In order to analytically investigate the change to the classical Freedericksz transition under an applied pressure gradient, we shall now  consider the behaviour of system (\ref{SS1}), (\ref{SS2}) for low  Ericksen numbers and close to the critical Freedericksz transition voltage so that the director angle may be assumed small. We make the further assumption that the velocity profile is symmetric with respect to the centre of the cell so that $v_y(0.5)=0$.  From (\ref{SS2}), we  can derive 
\begin{equation}v_y = \df{1}{2{g}(\theta)}(1-2y).\label{gradv}\end{equation}
Substituting (\ref{gradv}) into (\ref{SS1}) and assuming $\theta(y)$ is small, we find by expanding to third order in $\theta(y)$,
\begin{eqnarray}
&& \theta_{yy}+\left(k-1\right)\left( \theta^2 \theta_{yy}  +  \theta\,  (\theta_y)^{2}\right) 
 +  
 \pi^2 \bar{V}^2\theta\left(1 - \df{2}{3} \theta^2\right)   \nonumber\\
&&\mbox{\quad}
 \hspace{2cm}+ \df{  {\cal{E}} \mu ( 2y-1) }{2\gamma_1 \eta_{1}^2}\left(\alpha_{3}\eta_1-\theta^2 (\alpha_{2}\eta_{1}+\alpha_{3}\eta_{2}+\alpha_{3}\alpha_1)\right)  =0. \label{thtay}
 \end{eqnarray}
Considering the first two non-trivial modes ($n=1,\,2$), we seek solutions of \eqref{thtay} of the form
\begin{equation}\theta(y)= \cA \sin( \pi y) + \cB \sin(2\pi y),\label{thguess}\end{equation}
where constants $\cA$ and $\cB$ are to be determined. Clearly this form for $\theta(y)$ cannot satisfy \eqref{thtay} for all $y\in[0,\,1]$. However, using the standard approach of weakly nonlinear analysis, we substitute \eqref{thguess} into the differential equation (\ref{thtay}) and multiply the resulting equation by, in turn, $\sin(\pi y)$ and $\sin(2\pi y)$ before integrating across the cell from $y=0$ to $1$. This produces two simultaneous cubic polynomial equations to solve for $\cA$ and $\cB$,
\begin{eqnarray}
\cA\left((     \bar{V}^2   +   k  -1)
  \cA^2    +  \bigl(   2    \bar{V}^2 
   +5(k-1)    \bigr)   \cB^2   -  3 {\cal{E}}^* \cB  -2 ( \bar{V}^2-1)  \right)&=&0, \label{poly1}\\[3pt] 
 \cB\left(6\bigl(   \bar{V}^2
  +4(k-1) \bigr)    \cB^2+  6 \bigl(   2 \bar{V}^2  +5 (k-1)   \bigr )\cA^2  
    -  16{\cal{E}}^*\cB   \right.\mbox{\qquad}&&  \nonumber \\  \mbox{\quad} 
   \left.-12 ( \bar{V}^2    -4  )\right)-  9 {\cal{E}}^*\cA^2&=&-24\Delta{\cal{E}}^*, \label{poly2}
   \end{eqnarray}
where \begin{equation}\Delta=\df{\alpha_{{3}} \eta_1  }{\alpha_{{2}} \eta_1 + \alpha_3\eta_2 + \alpha_3\alpha_{1}    },\qquad   {\cal{E}}^*=\df{  \alpha_3\mu }{2\pi^3 \eta_1 \gamma_1\Delta}{\cal{E}},\label{est}\end{equation}
and we employ the  Miesowicz viscosities~\cite{IWS1}
$$\eta_1 = \frac{1}{2}(\alpha_3 + \alpha_4 + \alpha_6),\qquad \eta_2 = \frac{1}{2}(-\alpha_2 + \alpha_4 + \alpha_5).$$

By examining the parameter values for which real solutions of (\ref{poly1}) and (\ref{poly2}) can be found,  we can demarcate the $(\bar{V},\,{\cal{E}})$ space into regions corresponding to different modes in a manner similar to Fig.~\ref{fig7}. 
In order to find these roots, we form the resultant of the two cubic equations with respect to $\cA$ or $\cB$, a ninth-order polynomial, before determine regions where real roots for $\cA$ and $\cB$ can exist. Numerically it is difficult to produce results in terms all four parameters $\cal{E^*}$, $\bar{V}$, $k$ and $\Delta$ that appear in (\ref{poly1}) and (\ref{poly2}). However, we can obtain some information if we fix the value of $k=K_3/K_1$ and concentrate on the more important ratios involving viscosities, voltage and Ericksen number. For example,  if we concentrate on the region close to $\bar{V}=1$ and small Ericksen number for  the value of $k$ used earlier, we can  approximate the Freedericksz transition boundary as  \begin{equation}
 \bar{V}^2 \approx 1 + \Bigl( 1 + \df{2}{9}(5k-3)\Delta \Bigr) 
\Delta {\cal{E^*}}^2
+\df{1}{27}\Bigl( 1 
+ \df{2}{9}(19k+48)\Delta  + \df{8}{9}(k+3)(5k-3)\Delta^2        \Bigr)\Delta^2 {\cal{E^*}}^4
 \label{powapp}     \end{equation}
where $\cal{E}^*$ is the Ericksen number rescaled by ratios of viscosities introduced in (\ref{est}). 
\begin{figure}
 \begin{center} 
 \resizebox{.7\textwidth}{!}{\includegraphics{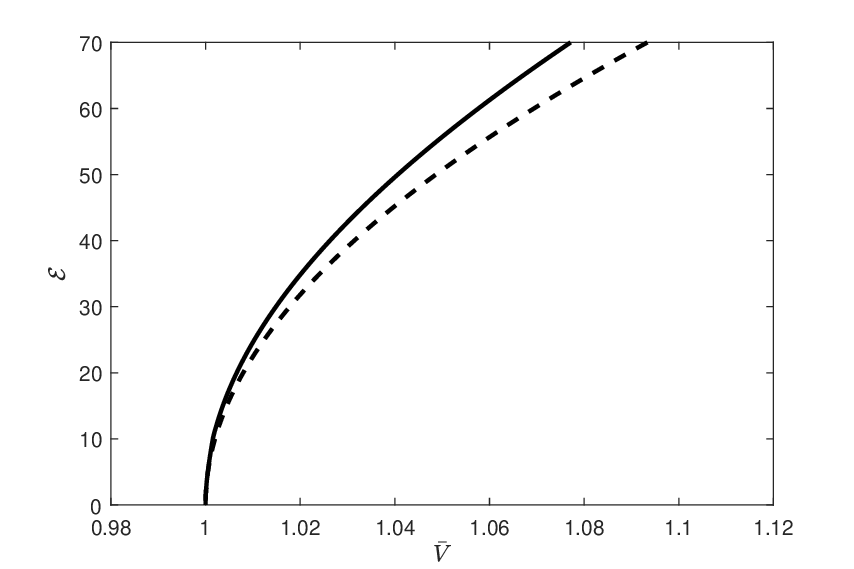}}
 \caption{\label{fig9}Analytic Freedericksz boundary (dashed curve) in normalised voltage-Ericksen number space, $(\bar{V},\,{{\cal E}})$, compared with  equivalent plot for the full numerical system from Fig.~\ref{fig7} (solid curve).   }  
 \end{center}
 \end{figure}
The curve obtained from (\ref{powapp}) is shown  in  Fig.~\ref{fig9} and compared with  the equivalent curve for the full system taken from the inset plot of Figure \ref{fig7}. 
While the curve given by eq.~\eqref{powapp} is clearly only a good approximation to the numerical form for  small Ericksen numbers, the analytic expression does   illustrate how the pressure-driven flow  delays the Freedericksz transition by increasing the critical voltage.


%

\section{Conclusions/Discussion}
In this paper, we have examined how pressure-driven flow modifies the classical Freedericksz transition in nematic liquid crystals, revealing a rich interplay between flow-induced and field-induced director reorientation. By employing the Ericksen-Leslie theoretical framework, we have shown that the presence of a pressure gradient alters the nature of the bifurcation at the Freedericksz transition, transforming it from supercritical to subcritical and shifting the critical voltage to higher values. This delay in the transition highlights the stabilising influence of flow on the initial director configuration.

As the pressure gradient increases, the system exhibits multistability, with multiple attracting steady states emerging over a broad range of voltages. Notably, for sufficiently high Ericksen numbers, multiple steady states exist even at zero applied voltage, including a novel flow-stabilised configuration that is not present in the absence of flow. These findings underscore the capacity of flow to promote new director states and reshape the bifurcation landscape of the system.

The introduction of flow also breaks the symmetry of the director profile, leading to asymmetric steady states and enabling hysteresis-like behavior in transitions between configurations. This symmetry breaking is particularly evident in the coexistence of symmetric and antisymmetric modes, and  we have suggested mechanisms to move between some of these states using voltage control, although switching to higher order modes may be difficult. A weakly nonlinear analysis near the classical Freedericksz transition point provides an analytical approximation for the delayed transition boundary in terms of geometric and material parameters, which aligns well with numerical results for small Ericksen numbers.

The presence of multiple attracting steady states in a real situation may also lead to spatial inhomogeneity of the director structure, with different regions of a device exhibiting different modes and defects in between. This situation could then lead to optical spatial inhomogeneity. However, if sufficiently controllable, these novel flow-electric field induced configurations give the possibility of director and flow hysteresis in microfluidic devices that could be used as switching mechanisms between multiple attracting states. Future investigations may explore the influence of anchoring conditions, confinement geometries, and dynamic switching protocols to further exploit the interplay between flow and field in nematic materials.

\vspace{0.5cm}
\noindent {\bf Statements and Declarations}

\noindent {\bf Data availability}: All methods to produce the data  of the paper are available within the paper itself.

\noindent {\bf Acknowledgements}: The authors are very grateful to Prof.~C.V.~Brown for useful discussions in the initial stages of this work. For the purpose of open access, the authors have applied a Creative Commons Attribution (CC-BY) licence to any Author Accepted Manuscript version arising from this submission. 

\noindent {\bf Author contributions}: Both authors conceived of the research and methodology, carried out the calculations and analysis, discussed the results and their interpretation, and contributed to the editing and final version of the manuscript.

\noindent {\bf Competing interests}: The authors declare no competing interests.

\end{document}